\documentclass[%
 reprint,
 aps,
 superscriptaddress,
 raggedbottom,
 prl]{revtex4-2}
\usepackage{CJK}

\usepackage{amsmath,amssymb,mathtools,extarrows,mathrsfs}
\usepackage{tikz,float,subcaption,graphbox}
\usetikzlibrary{matrix,arrows.meta,fit,positioning,decorations.markings,shapes.geometric}

\usepackage{titlesec}
\titleformat{\section}{\centering\normalsize\normalfont\bf}{\thesection}{0em}{}
\usepackage{hyperref}
\usepackage{comment}
\usepackage{tabularray}
\usepackage[T1]{fontenc}

\definecolor{lightblue}{rgb}{0.36, 0.51, 0.71}
\definecolor{lightyellow}{rgb}{0.88, 0.61, 0.14}
\definecolor{darkgreen}{rgb}{0.6, 0.6, 0.35}
\definecolor{lightred}{rgb}{0.94, 0.50, 0.50}

\def\ket#1{\langle #1 \rangle}
\begin{document}
\begin{CJK*}{UTF8}{gbsn}
\title{Heptagon Symbols at Five Loops and All-Loop Sequences}

\author{Song He (何颂)}\email{songhe@itp.ac.cn}\affiliation{Institute of Theoretical Physics, Chinese Academy of Sciences, Beijing 100190, China}\affiliation{School of Fundamental Physics and Mathematical Sciences, Hangzhou Institute for Advanced Study, Hangzhou, Zhejiang 310024, China}\affiliation{Peng Huanwu Center for Fundamental Theory, Hefei, Anhui 230026, P. R. China}
\author{Xuhang Jiang (姜旭航)}\email{xhjiang@itp.ac.cn}\affiliation{Institute of Theoretical Physics, Chinese Academy of Sciences, Beijing 100190, China}
\author{Xiang Li (李想)}\email{lixiang@itp.ac.cn}\affiliation{Institute of Theoretical Physics, Chinese Academy of Sciences, Beijing 100190, China}\affiliation{School of Physical Sciences, University of Chinese Academy of Sciences, Beijing 100049, China}
\author{Jiahao Liu (刘家昊)}\email{liujiahao@itp.ac.cn}\affiliation{Institute of Theoretical Physics, Chinese Academy of Sciences, Beijing 100190, China}\affiliation{School of Physical Sciences, University of Chinese Academy of Sciences, Beijing 100049, China}
\date{\today}

\begin{abstract}
We revisit the symbol bootstrap program for the seven-particle MHV and NMHV amplitudes in planar ${\cal N}=4$ super-Yang-Mills (SYM) based on the alphabet associated with the $E_6$ cluster algebra. After imposing integrability, cluster adjacency (or extended Steinmann), first- and last-entry conditions, the solution space is already highly restrictive: {\it e.g.} for MHV case there are exactly $1,1,2,3,4$ parity-invariant solutions for $L=1,2,\cdots, 5$, which automatically satisfy dihedral symmetry. Remarkably, after further requiring a well-defined collinear limit, we find a {\it unique} solution for both MHV and NMHV sectors where all coefficients ({\it e.g.} more than $3.1\times 10^{10}$ for MHV at $L=5$) turn out to be {\it integers}. Furthermore, we observe recurrent patterns for coefficients of special words in $E_6$ symbols mirroring those found for the $C_2$ symbol of three-point form factors, which lead to numerous predictions in the form of all-loop sequences. As an initial application, we show that these sequences uniquely fix the MHV symbol through five loops without input from collinear limits. Given the simplicity and surprisingly strong constraining power of both physical constraints and the newly observed sequences, we conjecture that there is a unique symbol satisfying these constraints at any loop order.  
\end{abstract}

\maketitle
\end{CJK*}

\section{Introduction}

Recent years have witnessed an explosion in our understanding of new structures of Quantum Field Theory, with the planar $\mathcal N{=}4$ super-Yang-Mills theory (SYM) serving as a particularly fertile testing ground. This theory has been explored through a variety of approaches including perturbative calculations, the AdS/CFT correspondence, and integrability, revealing a wealth of deep mathematical structures. For all-loop (gluon) scattering amplitudes in the theory, not only do we have a geometrical reformulation of its all-loop integrands via positive Grassmannian~\cite{Arkani-Hamed:2009ljj,Arkani-Hamed:2012zlh} and the amplituhedron~\cite{Arkani-Hamed:2013jha,Arkani-Hamed:2013kca}, but physical understandings and mathematical structures have also emerged for the amplitudes after loop integrations, which are dual to null polygonal super-Wilson loops~\cite{Alday:2007hr,Alday:2007he,Brandhuber:2007yx,Drummond:2007bm,Drummond:2007au,Bern:2008ap,Drummond:2007cf,Drummond:2008aq,Berkovits:2008ic,Mason:2010yk,Caron-Huot:2010ryg} and enjoy the famous dual conformal symmetry~\cite{Drummond:2006rz,Drummond:2008vq,Hodges:2009hk}. In particular, there has been a heroic bootstrap program for the $n=6,7$ amplitudes (dual to hexagon and heptagon) to impressively high loops~\footnote{The $n=4,5$ amplitudes are known to all loops by the famous Bern-Dixon-Smirnov ansatz, which is a consequence of Infrared exponentiation and dual conformal symmetry~\cite{Bern:2005iz}.}, which evaluate to multiple-polylogarithmic functions with an important coproduct structures~\cite{Goncharov:2005sla,Brown:2011ik,Duhr:2012fh} (loosely) known as the {\it symbol}~\cite{Goncharov:2009lql,Goncharov:2010jf}. The key input of the program is the well-tested conjecture that the {\it symbol alphabet} for $n=6,7$ consists of $9$ and $42$ dual-conformal-invariant (DCI) letters, which corresponds to finite cluster algebras associated with their kinematic space, $A_3 \sim Gr(4,6)$ and $E_6 \sim Gr(4,7)$ respectively (for $n>7$ such cluster algebras become infinite)~\cite{Golden:2013xva,Golden:2014xqa}. By imposing various physical constraints, the MHV and NMHV hexagons have been determined to very high loops~\cite{Dixon:2011nj,Dixon:2011pw,Dixon:2013eka,Dixon:2014iba,Dixon:2014voa,Dixon:2015iva,Caron-Huot:2016owq,Caron-Huot:2019vjl}. Intriguingly, on a parity-preserving surface where $A_3$
reduces to $C_2$, the MHV hexagon is antipodally dual to the three-gluon form factor (FF) of chiral stress-tensor multiplet~\cite{Dixon:2021tdw}, both of which have been bootstrapped up to eight loops~\cite{Dixon:2022rse,Dixon:2023kop}! 

The heptagon bootstrap based on $E_6$ alphabet has been carried out up to four loops~\cite{Dixon:2016nkn,Drummond:2018caf}, where the size of the ansatz hence the computational challenge grows more quickly than the heptagon/FF due to the considerably larger alphabet. However, as already observed in~\cite{Dixon:2016nkn}, the solution space actually seems much more restrictive: up to four loops there is a unique symbol after imposing conditions including first- and last-entry conditions, extended Steinmann~\cite{Caron-Huot:2019bsq} or the so-called ``cluster adjacency''~\cite{Drummond:2017ssj,Drummond:2018dfd}, symmetries, as well as well-defined collinear limits.  This is to be compared with the hexagon/FF where various more complicated conditions such as OPE expansions and multi-Regge limits are needed at relatively low loop orders~\cite{Dixon:2011pw,Dixon:2020bbt}. Note that the heptagon bootstrap contains hexagons in collinear limits, and via the antipodal duality, also the three-point tr$\phi^2$ FF, thus it would be very interesting to gather higher-loop data and to see if simple conditions still suffice. Moreover, the resulting $E_6$ symbols had not been studied from a more mathematical perspective: what is really special about these integrable, length-($2L$) $E_6$ symbols (besides satisfying basic conditions)? Very recently interesting recurrent features have been discovered for the $C_2$ symbols of the three-point tr$\phi^2$ FF which makes all-loop predictions for coefficients of certain words~\cite{Cai:2025atc}, and we find it very tempting to ask if similar sequences exist for the heptagons~\footnote{Note that on a parity-invariant subspace of MHV heptagon, the $E_6$ symbol reduces to a $F_4$ symbol (a folded cluster algebra), which in the collinear limit directly contains the $C_2$ symbol for the three-point FF (as the antipodal dual of the $C_2$ MHV hexagon).}. 

In this Letter, with the help of the powerful solver \texttt{SparseRREF}~\cite{SparseRREF}, we take a key step towards answering these questions: first we confirm that these physical conditions again uniquely determine the MHV and NMHV heptagon symbols up to five loops, which is a much more demanding computation than previous ones. The MHV symbol contains $31,199,389,998$ length-$10$ $E_6$ words, while for NMHV the three independent symbols contain $40,983,053,954$ ($E_{14}$),  $51,073,911,501$ ($E_{12}$),  $83,845,722,184$ ($E_0$) words respectively; remarkably all of their coefficients are {\it integers}, which exhibit extremely rich combinatorial structures! In particular, in both MHV and NMHV symbols we discover all-loop sequences for coefficients of special words with repeated last (or first) entries, which are inferred from lower-loop data and expected to hold for all loops. We show that these sequences also fix the MHV symbol through five loops without the need of collinear limits. Similar but slightly different sequences also exist for (the MHV and NMHV) hexagons and three-point FF (of tr$\phi^2$ and tr$\phi^3$).

\section{\texorpdfstring{$E_6$}{E6} bootstrap for the heptagon symbols}

We consider the planar limit $N_c\to \infty$ of ${\cal N}=4$ SYM theory with $g_{\rm YM}^2 N_c$ fixed, and the perturbative expansion of any quantity reads $F=\sum_{L=0}^\infty g^{2L} F^{(L)}$ with $g^2\equiv g_{\rm YM}^2 N_c/(16 \pi^2)$. The $n$-point scattering amplitudes or $n$-gon Wilson loops depend on dual coordinates defined via $p_a=x_{a{+}1}-x_a$ for $a=1,\dots, n$. There are different schemes for subtracting the all-loop exponentiated divergences~\cite{Bern:2005iz,Caron-Huot:2019bsq,Dixon:2016nkn,Caron-Huot:2020bkp}, and we focus on the BDS-like subtraction: for $n \, {\rm mod \, 4} \neq 0$, we define the $\text{N}^k\text{MHV}$ BDS-like normalized amplitudes, ${\cal E}_{n,k}$, which depends on $3(n{-}5)$ conformal invariant cross-ratios, via $\mathcal{A}_{n,k}={\cal E}_{n,k} \mathcal{A}_n^{\text{BDS-like}}$, where the explicit formula can be found in the appendix. It is convenient to introduce {\it momentum twistors}~\cite{Hodges:2009hk}, $Z_a^I=(\lambda_a^\alpha, x_a^{\alpha, \dot{\alpha}} \lambda^\alpha_a)$ with $SL(4)$ index $I:=(\alpha, \dot{\alpha})$, and the kinematic space in terms of them is equivalent to the Grassmannian mod torus action, $Gr(4,n)/T$, and conformal invariant quantities can be expressed as homogeneous function of {\it four-brackets} defined as $\langle i j k l\rangle:={\rm det} (Z_i Z_j Z_k Z_l)$; in particular $x_{i,j}^2:=(x_i-x_j)^2=(p_i + \cdots + p_{j{-}1})^2 \propto \langle i{-}1 i j{-}1 j\rangle$, where the proportionality factor drops out in conformal invariant combinations. In appendix, we provide more details of the kinematics and especially the parametrization of $n=7$ momentum twistors in terms of $3(n{-}5)=6$ independent variables. 

The $n\leq 7$ amplitudes at $L$ loops are expected to evaluate to multiple polylogarithms (MPL) of weight $w=2L$, which can be defined recursively by its derivatives, $\mathrm{d} F^{(w)}=\sum_{\{\phi_\alpha\}} F^{(w{-}1)}_{\phi_\alpha} \mathrm{d}\log \phi_\alpha$ with the $w=1$ case reduced to the usual logarithm whose derivatives involve rational numbers, $F^{(0)}_{\phi_\alpha}$. The collection of all arguments of $\mathrm{d}\log$, $\{\phi_\alpha\}$, is the {\it symbol alphabet}~\cite{Goncharov:2010jf}, which encodes positions of all possible branch points of $F^{(w)}$. There is a coaction acting on MPL functions (more details can be found in~\cite{Duhr:2012fh}), and the maximally iterated coproduct defines the {\it symbol}:
\begin{equation}
S[F^{(w)}]=
\sum_{\phi_{\alpha_1}, \dots, \phi{\alpha_w}} F^{(0)}_{\phi_{\alpha_1}, \dots, \phi_{\alpha_w}} [\mspace{1mu}\log \phi_{\alpha_1} \otimes \cdots \otimes \log \phi_{\alpha_w}]\,,   
\end{equation}
where $F^{(0)}$ are rational numbers, and we often drop the $\log$ and $\otimes$, and denote a {\it word} as $[\mspace{1mu}\phi_{\alpha_1}, \dots, \phi_{\alpha_w}]$. 

\paragraph{$E_6$ basic conditions}
A remarkable observation is that the {\it symbol alphabet} for $n=6,7$ is captured by the {\it cluster algebra} associated with $Gr(4,n)/T$, which are known as $A_3$- and $E_6$-type respectively~\cite{Golden:2013xva}. Concretely, the heptagon alphabet $\{a_\alpha\}$ consists of $42$ conformally invariant combination of {\it cluster ${\cal A}$-coordinates} of $Gr(4,7)\sim E_6$, which can be chosen canonically as $42$ {\it homogeneous} ${\cal A}$-coordinates, grouped into six cyclic orbits as $a_{i j}$~\cite{Drummond:2014ffa} for $i=1,\dots, 6$ and $j=1,\dots, 7$:
\begin{align}
\label{Salphabet}
a_{1 1} &= \frac{\ket{1234}\ket{1567}{\ket{2367}}}{\ket{1237}\ket{1267}\ket{3456}}\,, \quad & 
a_{6 1} &= \frac{{\ket{1(34)(56)(72)}}}{\ket{1234}\ket{1567}}\,,\nonumber\\ 
a_{2 1} &= \frac{\ket{1234}{\ket{2567}}}{\ket{1267}\ket{2345}}\,, \quad &
a_{3 1} &= \frac{\ket{1567}{\ket{2347}}}{\ket{1237}\ket{4567}}\, , \nonumber\\
a_{4 1} &= \frac{{\ket{2457}}\ket{3456}}{\ket{2345}\ket{4567}}\,, \quad &
a_{5 1} &= \frac{{\ket{1(23)(45)(67)}}}{\ket{1234}\ket{1567}}\,,
\end{align}
which we also denote as $W_{1,\dots, 42}=\{a_{1,1},a_{1,2},\ldots,a_{6,7}\}$. The dihedral symmetry of the heptagon acts nicely on these letters: we have $a_{i,j}\to a_{i,j{+}1}$ ($i=1,\cdots, 6$) for cyclic transformation, and $a_{2,j}\leftrightarrow a_{3,8-j}$, $a_{i,j}\to a_{i,8-j}$ for $i\neq 2,3$ under  the flip. The main question is then how to constrain the rational coefficients of all length-$(2L)$ words for the symbols of MHV and (independent functions of) NMHV amplitudes:
\begin{equation}
\sum_{\{\alpha\}} c_{\{\alpha\}} [\mspace{1mu}a_{\alpha_1}, \dots, a_{\alpha_{2L}}]\,,
\end{equation}
where $\{a_\alpha\}$ denotes the heptagon symbol alphabet.

The most basic conditions come from the requirement that the symbol must be {\it integrable}, {\it i.e.} it is the symbol of an actual function, which must have commuting second derivatives w.r.t. any two independent variables. These imply the well-known {\it integrability conditions}, which are linear relations on the double coproducts for two adjacent entries of the symbol. Remarkably, as we review in the appendix, such conditions have a natural interpretation from the cluster algebra, which also gives an invariant formulation without referring to any explicit parametrization: they can be generated from identities among $\mathrm{d}\log$ two-forms associated with a cluster under mutations. By walking through all $833$ clusters, we generate all identities among $\mathrm{d}\log$ two-forms, which provide $729$ independent linear constraints on the naive ${42 \choose 2}=861$ two-forms. 

Similar to but simpler than integrability, {\it cluster adjacency}, which is equivalent to extended Steinmann relations~\footnote{It has been shown in \cite{Golden:2019kks} that cluster adjacency implies the extended Steinmann relations at all multiplicities. Especially for heptagon amplitudes, they are observed to be equivalent to each other~\cite{Drummond:2017ssj}.}, states that on any two adjacent slots, the pair of letters must belong to a cluster, and it is straightforward to enumerate all $840$ pairs as done in~\cite{Drummond:2017ssj}. Altogether, they provide linear constraints on any adjacent slots. Furthermore, the {\it first-entry} conditions~\cite{Gaiotto:2011dt}, which follow from the physical discontinuity by considering the unitary cuts, require that $a_{\alpha_1} \in \{a_{1,j={1,2,\cdots, 7}}\}=\{W_{1,2,...7}\}$. We emphasize that these conditions are rather universal, which apply to both MHV and NMHV amplitudes. The difference is that they obey different last entries, and they have different symmetry. We will state those conditions for MHV and NMHV, respectively. 

\paragraph{MHV}

For the symbol of MHV heptagon ${\cal S}^{(L)}:=S[{\cal E}_{7,0}^{(L)}]$, last-entry conditions (derived from the $\bar{Q}$ equations~\cite{Caron-Huot:2011dec}) state that only the 14 letters $a_{2j}$ and $a_{3j}$ appear in the last entry, $a_{\alpha_{2L}}\in \{W_{8, \cdots, 21}\}$~\footnote{Note that the first entries correspond to subalgebras, or {\it facets} of the $E_6$ polytope, of type $A_5$, and the last entries correspond to subalgebras of type $D_5$, both of which are quite special in the Dynkin diagram/polytope.}. As we have checked up to $L=5$, once we impose these conditions above, dihedral symmetry remarkably becomes automatic. We also impose parity symmetry, which slightly reduces the ansatz starting from $L=3$. As shown in Table~\ref{tab:counting}, after imposing last-entry and parity (dihedral symmetry is automatic), we find exactly $1,1,2,3,4$ solutions for $L=1,2,3,4,5$, which shows how restrictively these conditions are for the MHV heptagon. 

\begin{table}[htb]
    \centering
    \begin{tblr}{colsep=4pt,belowsep=0.7pt,colspec={Q[c,colsep=6pt]|[dotted]Q[c]|[dotted]Q[c]|[dotted]Q[c]|[dotted]Q[c]|[dotted]Q[c]},rowspec={|[1pt]Q[m]|Q[m]|[dotted]Q[m]|Q[m]|[dotted]Q[m]|Q[m]|Q[m]|[1pt]}}
        $L$ & 1 & 2 & 3 & 4 & 5  \\
        \SetCell[r=2]{c}{initial} & {\color{lightred}28} & {\color{lightred}308} & {\color{lightred}2555} & {\color{lightred}17471} & {\color{lightred}?} \\
        & {\color{lightblue}882} & {\color{lightblue}12222} & {\color{lightblue}114786} & {\color{lightblue}356076} & {\color{lightblue}5415480} \\
        \SetCell[r=2]{c}{$+$ last $+$ sym.}& {\color{lightred}1}  & {\color{lightred}1}  & {\color{lightred}2}  & {\color{lightred}3}  & {\color{lightred}4} \\
        & {\color{lightblue}5}  & {\color{lightblue}11}  & {\color{lightblue}24}  & {\color{lightblue}56}  & {\color{lightblue}136} \\
        $+$ collinear & 1 & 1 & 1 & 1 & 1  \\
        or $+$ $\rm MHV_{seq.}$ & {\color{lightred}0} & {\color{lightred}0} & {\color{lightred}0} & {\color{lightred}0} & {\color{lightred}0} \\
    \end{tblr}
    \caption{Numbers of parameters in the symbol bootstrap for {\color{lightred} MHV} and {\color{lightblue} NMHV} up to $L=5$: first in the ``initial'' ansatz (with first-entry conditions, $E_6$ integrability and adjacency up to weight-$2L$ for {\color{lightred} MHV} and up to weight-$(2L{-}1)$ for {\color{lightblue} NMHV}), then after imposing last-entry (and adjacency+integrability for the last-two entry of {\color{lightblue} NMHV}) and symmetries; finally, imposing that the collinear limit exists fixes them up to an overall normalization; in the {\color{lightred} MHV} case alternatively the sequences fix everything.}
    \label{tab:counting}
\end{table}
\paragraph{NMHV} The NMHV amplitude is given as a linear combination of R-invariants~\cite{Drummond:2008vq,Drummond:2008bq,Mason:2009qx}, which encode helicity information via Grassmann variables, and for heptagon they can be denoted as $(12):=[34567]$, $(13):=[24567]$, $(14):=[23567]$, and their cyclic images; these $3\times 7$ R-invariants are not linearly independent, and a nice basis can be chosen as $(12)$, $(14)$, plus cyclic, together with the tree amplitude $\mathcal{A}_7^{(0)} = \frac{3}{7} \, (12) + \frac{1}{7} \, (13) + \frac{2}{7} \, (14) ~ + \text{cyclic}$. Therefore, the BDS-like heptagon NMHV amplitudes can be written in terms of $15$ transcendental functions:
\begin{equation}
    \mathcal{E}_{7,1} = \mathcal{A}_7^{(0)} \, E_0 + \big[ (12) \, E_{12} + (14) \, E_{14} + \text{ cyclic} \big],
\end{equation}
whose seeds are $E_{12}$, $E_{14}$ and $E_0$, and the goal is to bootstrap the $E_6$ symbol of these weight-($2L$) MPL functions at $L$ loops. Similar to MHV case, they are expected to satisfy the $E_6$ basic conditions, but the last-entry conditions differ. The latter has been derived from $\bar{Q}$ equations in~\cite{Dixon:2016nkn}. It is more convenient to impose such conditions on the redundant basis, where the last entries are connected to their accompanying R-invariants, which we denote as
\begin{equation}\label{eq:ansatz}
      S \left[\mathcal{E}_{7,1}^{(L)} \right] = e_{12}^{(L)} \, (12) + e_{13}^{(L)} \, (13) + e_{14}^{(L)} \, (14) + \text{cyclic}.
\end{equation}
The $e_{ij}^{(L)}$ are tensor products of the form
\begin{equation}\label{eq:eij_coproduct}
    e_{ij}^{(L)} = \sum_{\phi_\alpha\, \in\, \text{hns}_{\bar{Q}}[(ij)]}\,\, \sum_{k} \,\, c^{(ij)}_{k, \alpha}\,\, f^{(2L-1)}_{k}  \otimes \phi_{\alpha}\,,
\end{equation}
where $f^{(2L-1)}_{k}$ are symbols that satisfy the $E_6$ basic conditions at weight-$(2L-1)$, and $\text{hns}_{\bar{Q}}[(ij)]$ denote the homogeneous neighbor sets of R-invariant $(ij)$ compatible with $\bar{Q}$ equation~\cite{Drummond:2018dfd,Drummond:2018caf}:
\begin{equation}
\begin{aligned}
    \text{hns}_{\bar{Q}}[(12)] &= \{ a_{15}, a_{21}, a_{26}, a_{32}, a_{34}, a_{53}, a_{57}\} \, ,\\
    \text{hns}_{\bar{Q}}[(13)] &= \{ a_{21}, a_{23}, a_{31}, a_{33}, a_{41}, a_{43}, a_{62}\} \, ,\\
    \text{hns}_{\bar{Q}}[(14)] &= \{ a_{11}, a_{14}, a_{21}, a_{24}, a_{31}, a_{34}, a_{46}\} \, , \\
    & \hspace{11em}\text{\& cyclic}\,.
\end{aligned}
\end{equation}
Note that we do not impose integrability on the last two ($(2L-1)^{\text{th}}$ and $(2L)^{\text{th}}$) slots of $e_{ij}^{(L)}$, and they only become integrable there when combined to the independent $E$-functions:
\begin{equation}
\begin{aligned}\label{eq:eij_to_Eij}
    & S \left[ E^{(L)}_0 \right]  = \sum_{i=1}^{7} e^{(L)}_{i\,i+2},  \,\, S\left[ E^{(L)}_{14} \right] = e^{(L)}_{14} - e^{(L)}_{16} - e^{(L)}_{46},\\
    & S \left[ E^{(L)}_{12} \right]  = e^{(L)}_{12}- e^{(L)}_{16} - e^{(L)}_{24} - e^{(L)}_{46}\, .
\end{aligned}
\end{equation}
In Table~\ref{tab:counting}, we also present the numbers of parameters for NMHV bootstrap, where the initial ansatz for $e_{i,j}$ has $3\times 42$ times the number at weight $2L{-}1$; then we impose the last-entry conditions, $(2L{-}1, 2L)$-adjacency and $\mathbb{Z}_2$ symmetry of $e_{12}^{(L)},e_{13}^{(L)},e_{14}^{(L)}$, together with $(2L{-}1, 2L)$-integrability for $E^{(L)}_{0},E^{(L)}_{14}$, and $E^{(L)}_{12}$. Although not as restricted as MHV case, we see that this step still consists of a huge reduction in the solution space from the initial ansatz, {\it e.g.} from over 5.4 millions to $136$ at $L=5$. Another interesting observation is that through five loops, the integrability of $E_0$ does not lead to any new constraints, and all the parameters in it are already fixed by those in $E_{12}$ and $E_{14}$ at this stage. 

\section{Collinear limits and the unique symbols}
Finally, we impose the existence of collinear limits for MHV and NMHV heptagons. We will take the same collinear limit $7\parallel 6$ as in \cite{Drummond:2014ffa,Dixon:2016nkn}.
\begin{equation}\label{eq:momtwicoll}
    Z_7\to Z_{6}+\epsilon \frac{\langle1246\rangle}{\langle1245\rangle}Z_{5}+\epsilon\tau\frac{\langle2456\rangle}{\langle 1245\rangle}Z_{1}+\eta\frac{\langle1456\rangle}{\langle1245\rangle}Z_{2},
\end{equation}
where one sends $\eta \rightarrow 0$ and then $\epsilon \rightarrow 0$ while leaving $\tau$ fixed. There are totally 18 independent letters after taking the collinear limit. Nine of them are spurious letters which must cancel in the physical results. The remaining nine are symbol letters of six-point amplitudes $\{u,v,w,1-u,1-v,1-w,y_u,y_v,y_w\}$ \cite{Dixon:2011pw}. We find that seven of the nine spurious letters directly do not appear in our solution space. The remaining two spurious letters are $\eta$ and $\tau$. After redefining the remaining letters as follows,
\begin{equation}\label{eq:collinearbasis}
\begin{aligned}
    &\eta^{\prime}=\frac{\eta}{a},\, \tau^{\prime}=\tau \frac{xf}{d},\,a=\sqrt{\frac{u}{vw}},\, b=\sqrt{\frac{v}{uw}},\, c=\sqrt{\frac{w}{uv}}, \\
    &d=\sqrt{\frac{(1-u)(1-v)}{uv}}, \, e=\sqrt{\frac{(1-v)(1-w)}{vw}}, \,\\
    &f=\sqrt{\frac{(1-u)(1-w)}{uw}}, \, x=\sqrt{y_{u}},\, y=\sqrt{y_{v}},\, z=\sqrt{y_{w}},
\end{aligned}
\end{equation}
the collinear limit of $\mathcal{S}^{(1)}$ takes the following simple form
\begin{equation}\label{eq:Y7simplified}
\begin{aligned}
    \mathcal{S}^{(1)} \stackrel{7 \parallel 6}{\longrightarrow} -2\left([\eta^{\prime},\eta^{\prime}]+[\tau^{\prime},\tau^{\prime}]+[a,e]+[b,f]+[c,d]\right).
\end{aligned}
\end{equation}
Note that it is the BDS-normalized amplitudes that has the well-defined collinear limit. For the BDS-like-normalized amplitudes, we should consider products of lower loop results. See the appendix for more detail.

For MHV amplitudes, the requirement of cancellation of $\eta^{\prime}$ and $\tau^{\prime}$ for BDS-normalized amplitudes $R_7^{(L)}$ will directly fix all the free parameters in the solution space. The transformation between BDS-like-normalized amplitudes and BDS-normalized amplitudes needs the calculation of powers of $\mathcal{E}^{(1)}_{7,0}$. For example, two-loop BDS-normalized amplitudes $R_{7}^{(2)}=\mathcal{E}^{(2)}_{7,0}-\frac{1}{2}\big(\mathcal{E}^{(1)}_{7,0}\big)^2$. All the dependence on $\eta^{\prime}$ and $\tau^{\prime}$ in the collinear limit of BDS-like-normalized amplitudes originates from powers of $\mathcal{E}^{(1)}_{7}$ and the use of \eqref{eq:Y7simplified} greatly simplifies the calculation. 

For NMHV amplitudes, we impose two conditions in the collinear limit. The first condition is the same as in \cite{Drummond:2018caf}: $E_{0}+E_{23}+E_{34}-\mathcal{E}_{7,0}$ vanishes in the collinear limits. This condition will fix 130 in all 136 free parameters. The second condition is that $E_{47}-E_{67}$ does not depend on the spurious letters $\eta^{\prime}$ and $\tau^{\prime}$ in the collinear limit \cite{Dixon:2016nkn}. This is similar to the MHV case. It fixes the remaining 6 parameters. As shown in Table.\ref{tab:counting}, the existence of collinear limit alone is powerful enough to fix the physical results up to five loops for both the seven-point MHV and NMHV amplitudes (one can also fix the overall normalization easily by comparing with the BDS-like normalization). It is natural to conjecture that this remains to be true to any loop order. 

\section{Recurrent sequences for the integer coefficients}
Let us first comment on some statistics about these $E_6$ symbols. For the MHV symbol, the number of words grows quickly as $42, 2184, 467250, 105403942$ and $31199389998$ for $L=1,\dots, 5$, but the ratio of this number to either that of all possible words ($42^{2L}$) or that of the so-called ``adjacency-allowed words'' is tiny and decreases with $L$; the latter counts the number of words satisfying cluster-adjacency, first- and last-entry conditions, which can be easily obtained (and by itself it is a tiny fraction of $42^{2L}$). We find that the ratio of the number of terms to that of ``adjacency-allowed words'' is $0.6,0.077,0.037,0.019,0.012$ up to $L=5$. This means that there must be a huge number of derived zeros (from integrability and further conditions such as collinear limits) which states that the majority of allowed words are in fact forbidden, similar to what was understood for the FF case~\cite{Cai:2025atc}. The counting for NMHV symbol and the ratio to that of allowed words are of the same magnitude as those for MHV.

Note that the overall normalization of both MHV and NMHV symbols can be fixed by collinear limits (eventually coming from the one-loop BDS-like functions), thus it is a rather non-trivial and remarkable fact that up to $L=5$, all of their coefficients are integers! We note that for MHV the largest coefficients (in magnitude) are associated with special words with the most repeated first entries, namely $[a_{1 1}^{2L{-}1},x]$ for $x=a_{2 1}, a_{2 2}$ or $a_{2 5}$, as well as dihedral images (in total $3\times 14$ terms); in the shorthand notation, these coefficients are 
\begin{equation}
    c_{1^{2L{-}1}, 8}=c_{1^{2L{-}1}, 9}=-c_{1^{2L{-}1}, 12}=(-)^L 4^{L{-}1} (2L{-}3)!!   
\end{equation}
which are $\pm 1, \pm 4, \pm 48, \pm 960, \pm 26880$ for up to $L=5$ (for $L=1$, they give the full answer: $S[{\cal E}_{7,0}^{(1)}]=-[a_{1 1}, \frac{a_{2 1} a_{3 7}}{a_{2 5}}] + {\rm dih.}$). This is the first all-loop sequence we observe, and it is equivalent to the recursion $c_{1^{2L{-}1}, x}=-4(2L{-}3) c_{1^{2L{-}3}, x}$ for $x=8, 9$ or $12$. We provide data and statistics for all coefficients in an ancillary file. 

Next, we consider coefficients for special words similar to those found for the $C_2$ symbols of three-point tr$\phi^2$ FF~\cite{Cai:2025atc}, namely a sub-word of length $m$, denoted as $X_m$, followed by $(2L{-}m)$ repeated last entries. For MHV heptagon, dihedral symmetry means that we only need to consider one repeated last entry, say $a_{2 1}$, {\it i.e.} we are interested in coefficients of $S[X_m, a_{2 1}^{2L{-}m}]$. The simplest case is for $m=1$, {\it e.g.} $S[a_{1 1}, a_{2 1}^{2L{-}1}]$, whose coefficient reads $c_{1,8^{2L{-}1}}=(-)^{L} 2^{L{-}1} (2L{-}3)!!=-1, 2, -12, 120, -1680$ for $L=1,\dots, 5$ (which is $1/2^{L{-}1}$ times $c_{1^{2L{-}1}, 8}$) and they satisfy the recursion $c_{1, 8^{2L{-}1}}=-2(2L{-}3) c_{1, 8^{2L{-}3}}$ instead. In general, we find the coefficients for such special words with $m\leq L$ take the form
\begin{equation}\label{eq:coeffpat}
\begin{aligned}
    c_{X_m, a_{2 1}^{2L{-}m}}=\,&(-)^{L{-}1} \, 2^{L{-}\lceil m/2\rceil}\, p_{X_m} (L) \times \\[-1pt] &  \times \left(2L{-}2 \lceil m/2\rceil -1\right)!!\, ,
\end{aligned}
\end{equation}
where $p_{X_m} (L)$ is a polynomial of degree $(\lceil \frac{m}{2} \rceil-1)$ in $L$ (depending on the sub-word $X_m$), that is degree $0$ for $m=1,2$ as discussed above, degree $1$ for $m=3,4$ and degree $2$ for $m=5,6$, {\it etc.} There are $25$ possible sub-words $X_2$ whose coefficients are of the form $c_{X_2,8^{2L{-}2}} = c_{X_2} (-)^{L} 2^{L{-}1} (2L{-}3)!!$ with $c_{X_2}=\pm 1, \pm 1/2$ depending on the sub-word and $338$ $X_3$ with degree $1$ polynomials. Note that a special word with $2L-m$ repeated last entries first appear at $L=\lfloor  m/2\rfloor +1$ loops, and we need $\lceil \frac{m}{2} \rceil$ data points to deduce $p_{X_m}(L)$. Previous data (up to four loops) can be used to verify the coefficients for $m=1,2,3$ and predict the $m=4$ coefficient at five loops. 

As an application, we use such sequences to constrain the solution space satisfying the basic conditions, and found that for $L=3,4,5$, the remaining $2,3,4$ unknowns are fixed by sequences extracted from lower loops~\footnote{This is highly non-trivial since the majority of the sequences serve as consistency checks with other conditions; for $L=1,2$, they fix the overall normalization.}. Remarkably, all the $4806$ constraints from $c_{X_4, a_{2 1}^{2L{-}4}}$ for $L=5$ can be resolved without contradiction and are in full agreement with all other sequences. From $L=5$ data, we obtain $67722$ new all-loop sequences for sub-words with $m=5$ (with $p_{X_5}(L)$ of degree $2$), and we record all sequences with $m\leq 5$ in an ancillary file.

For NMHV heptagon, we consider all possible repeated last entries for $E_0, E_{12}, E_{14}$, and while most of them still exhibit the same pattern as for MHV (up to $L=5$), we find that with repeated $a=a_{6 1}$ in $E_0$, $a=a_{6 3}$ or $a_{6 5}$ in $E_{12}$, $a=a_{6 5}$ in $E_{1,4}$, the coefficient becomes  
\begin{equation}\label{eq:NMHVpattern}
\begin{aligned}
    c_{X_m, a^{2L{-}m}}=\,&(-)^{L{-}1} \, (12)^{L{-}\lceil m/2\rceil}\, p_{X_m} (L) \times \\[-1pt] &  \times \left(2L{-}2 \lceil m/2\rceil -1\right)!!\, ,
\end{aligned}
\end{equation}
where $p_{X_m}(L)$ is a degree-$(\lceil \frac{m}{2} \rceil-1)$ polynomial in $L$. It turns out that the largest coefficients for these functions appear for the $m=1$ case where $p=\pm 1$, which, up to a sign, is
$(12)^{L{-}1} (2L{-}3)!!$ (or $3^{L{-}1}$ times the largest coefficients in the MHV sector)~\footnote{In both MHV and NMHV sectors, we also find (slightly more involved) sequences for words with $2L{-}m$ repeated first entries followed by sub-words $X_m$.}. We also record all available NMHV sequences in an ancillary file. 

\section{Conclusions and Outlook}

In this Letter, we restart the symbol bootstrap program for MHV and NMHV heptagon (seven-gluon amplitudes in ${\cal N}=4$ SYM), which can be viewed as a mathematical problem of finding length-$(2L)$ integrable $E_6$ symbols satisfying the basic conditions, and additional physical constraints such as collinear limits; indeed these suffice to fix a unique solution for both MHV and NMHV symbols up to $L=5$, where all coefficients turn out to be {\it integers} with remarkable patterns. In particular, the MHV heptagons through five loops can also be uniquely determined simply by such patterns inferred from lower-loop data, which are trivial to impose. 

There are several avenues for further investigations based on our results. It would be interesting to uplift five-loop results to function level~\cite{Dixon:2020cnr}, and study various limits such as multi-Regge kinematics~\cite{Bartels:2008ce,Bartels:2009vkz,Lipatov:2010ad,Bartels:2010tx,Dixon:2011pw,DelDuca:2016lad, Dixon:2021nzr}, OPE expansions~\cite{Alday:2010ku,Gaiotto:2010fk,Gaiotto:2011dt,Sever:2011da,Basso:2013vsa,Basso:2013aha,Basso:2014hfa}, which provide both further checks and invaluable data for future studies; it would also be interesting to study the double-spacelike collinear limits of~\cite{Duhr:2025lyg}. It becomes computationally rather challenging to move to six loops, but perhaps there is a better method which does not require first generating a huge ansatz but rather considered as ``optimizing'' conditions such as integrability {\it etc.} in the space of all allowed $E_6$ words.

We suspect that either the collinear limit or the sequences uniquely determines heptagon symbols to all loops, but for the latter we still need to better understand all possible sequences especially in the NMHV case (perhaps with the help of some machine-learning analysis~\cite{Cai:2024znx}). It would be highly desirable to see if the solution space of $E_6$ bootstrap is always so restrictive, and if all coefficients are integers with rich combinatorial structures~\footnote{It might be an even more universal phenomenon, {\it e.g.} the distribution of coefficients here resembles that of the coefficients for the $f$-graphs for planar integrand of four-point correlator in ${\cal N}=4$ SYM, known up to twelve loops~\cite{He:2024cej, Bourjaily:2025iad}}. As we will elaborate elsewhere, such all-loop sequences/recursions exist universally for a web of physical quantities (related to each other via collinear limits and antipodal duality): heptagons (MHV and NMHV), hexagons (MHV and NMHV)~\cite{Dixon:2023kop, Caron-Huot:2019vjl}, and three-point FF (with tr$\phi^2$ and tr$\phi^3$ operators)~\cite{Cai:2025atc, Dixon:2022rse, Basso:2024hlx}, thus there must be some common origin for their universal behaviors~\footnote{Relatedly, we would like to better understand the alphabet and symbols of four-point FF, which enjoys antipodal self-duality~\cite{Dixon:2022xqh, Dixon:2024yvq}, and even higher-point ones~\cite{Li:2024rkq}.}. It is natural to wonder, if we view the symbol of such a quantity as a map from possible length-$(2L)$ ``cluster words'' to integers, what could be the (combinatorial, cluster-algebraic and even geometrical) question whose answer gives these multi-loop scattering amplitudes/Wilson loops and form factors?

\section*{Acknowledgments}It is our pleasure to thank Lance Dixon and Zhenjie Li for numerous inspiring discussions and comments on the draft. SH has been supported by the National Natural Science Foundation of China under Grant No. 12225510, 12047503, 12247103, and by the New Cornerstone Science Foundation through the XPLORER PRIZE. Some of the results are obtained on the HPC cluster of ITP-CAS.

\bibliographystyle{physics}
\bibliography{reference.bib}

\appendix
\widetext
\section{Review of basic conditions for heptagon bootstrap}
The symbol alphabet of seven-particle amplitudes in $\mathcal{N}=4$ SYM can be described by cluster algebras associated with Grassmannian $\mathrm{Gr}(4,7)$, which is naturally related to the planar kinematics of seven massless particles in momentum twistor parameterization. Let us first briefly review the momentum twistor and the expressions of 42 symbol letters.  The kinematics of seven massless particles can be equally described by a heptagon in dual space, where the dual coordinates $x_{i}^{\mu}$ are related to momenta via $p_i^{\mu}=x_{i}^{\mu}-x_{i+1}^{\mu}$, $x_{i,i+1}^2=p_{i}^{2}=0$. The four-dimensional vector $x_{i}^{\mu}$ can be embedded in a six-dimensional projective space as a null vector, which is further written as an antisymmetric tensor of momentum twistors $(Z_{i-1},Z_{i})$. $Z_{i=1,\ldots,7}^{a}$ with $a=1,2,3,4$ are called momentum twistor variables and $Z_{i}\sim t_{i}Z_{i}$. They expand the Grassmannian $\mathrm{Gr}(4,7)$. In momentum twistor parameterization, we have
\begin{equation}\label{x2p}
    s_{i,\dots,j-1} = (p_{i} + p_{i+1} + \dots + p_{j-1})^{2} = x_{ij}^{2} = \frac{\ket{i{-}1\, i\, j{-}1\, j}}{\ket{i{-}1\, i}\ket{j{-}1\, j}}\,,
\end{equation}
where $\ket{i{-}1\,i\,j{-}1\,j}\equiv \ket{Z_{i{-}1}\,Z_{i}\,Z_{j{-}1}\,Z_{j}}$ and $\ket{i{-}1\, i}\equiv \ket{Z_{i{-}1}\,Z_{i}\,I_{\infty}}$, $I_{\infty}$ is the infinity line. However, physical quantities enjoy dual conformal symmetry, indicating that $I_{\infty}$ does not appear in the final results. The symbol alphabets can then be written as a set of dual conformal invariant (DCI) cross-ratios~\cite{Drummond:2014ffa}
\begin{align}\label{SalphabetAppx}
    a_{11} &= \frac{\ket{1234}\ket{1567}{\color{blue}\ket{2367}}}{\ket{1237}\ket{1267}\ket{3456}}\,, \quad & 
    a_{21} &= \frac{\ket{1234}{\color{blue}\ket{2567}}}{\ket{1267}\ket{2345}}\,, \quad &
    a_{31} &= \frac{\ket{1567}{\color{blue}\ket{2347}}}{\ket{1237}\ket{4567}}\, , \nonumber\\
    a_{41} &= \frac{{\color{blue}\ket{2457}}\ket{3456}}{\ket{2345}\ket{4567}}\,,\quad & 
    a_{51} &= \frac{{\color{blue}\ket{1(23)(45)(67)}}}{\ket{1234}\ket{1567}}\,, \quad &
    a_{61} &= \frac{{\color{blue}\ket{1(34)(56)(72)}}}{\ket{1234}\ket{1567}}\, 
\end{align}
and their cyclic images $a_{ij},j=2,\ldots,7$. Here $\ket{a(bc)(de)(fg)}=\ket{abcf}\ket{adeg}-\ket{abcg}\ket{adef}$. The symbol letters above can be described by the cluster algebra of $\mathrm{Gr}(4,7)$. We start from the initial quiver of $\mathrm{Gr}(4,7)$~\cite{Drummond:2017ssj}:
\begin{equation}
    \begin{tikzpicture}[>=stealth,boxed/.style={draw, rectangle}, every path/.style={shorten >=2pt, shorten <=2pt}]
    \matrix (m) [matrix of nodes,
        row sep=1.2em,
        column sep=2em,
        nodes={inner sep=1.5pt}] 
    {|[boxed]| $\ket{1234}$ &            &            &             \\
             & $\ket{1235}$   & $\ket{1236}$   & |[boxed]|$\ket{1237}$   \\
             & $\ket{1245}$   & $\ket{1256}$   & |[boxed]| $\ket{1267}$   \\
             & $\ket{1345}$   & $\ket{1456}$   & |[boxed]| $\ket{1567}$   \\
             & |[boxed]| $\ket{2345}$   & |[boxed]|$\ket{3456}$   & |[boxed]| $\ket{4567}$   \\
    };

    \draw[->] (m-1-1) -- (m-2-2);

    \draw[->] (m-2-2) -- (m-2-3);
    \draw[->] (m-2-3) -- (m-3-3);

    \draw[->] (m-2-2) -- (m-3-2);
    \draw[->] (m-2-3) -- (m-3-3);
    \draw[->] (m-2-3) -- (m-2-4);

    \draw[->] (m-3-2) -- (m-4-2);
    \draw[->] (m-3-2) -- (m-3-3);
    \draw[->] (m-3-3) -- (m-2-2);
    \draw[->] (m-3-3) -- (m-3-4);
    \draw[->] (m-3-3) -- (m-4-3);
    \draw[->] (m-3-4) -- (m-2-3);

    \draw[->] (m-4-2) -- (m-4-3);
    \draw[->] (m-4-2) -- (m-5-2);
    \draw[->] (m-4-3) -- (m-5-3);
    \draw[->] (m-4-3) -- (m-3-2);
    \draw[->] (m-4-3) -- (m-4-4);
    \draw[->] (m-4-4) -- (m-3-3);

    \draw[->] (m-5-4) -- (m-4-3);
\end{tikzpicture}
\end{equation}
Each node is assigned an $\mathcal{A}$ coordinate. There are seven boxed nodes with $\mathcal{A}$ coordinates $\ket{i\,i{+}1\,i{+}2\,i{+}3}$. They are called frozen nodes, which are not mutable. The rest nodes can be mutated according to a set of rules. The quiver can be described by the exchange matrix $[b_{ij}]$ which is defined by
\begin{equation}
    b_{ij}=\left\{
    \begin{array}{cc}
        l   & l\text{ arrows from node $i$ to node $j$} \\
        -l   & l\text{ arrows from node $j$ to node $i$} \\
        0 & \text{otherwise}
    \end{array}\right. .
\end{equation}
and if we mutate node $k$, the matrix $[b_{ij}]$ and the $\mathcal{A}$ coordinate $x_{k}$ on node $k$ will change (which is also called \textit{exchange relations}) as follows
\begin{equation}\label{eq:exchangerels}
    b_{ij}^{\prime} = \left\{
    \begin{array}{cc}
        -b_{ij} & k\in\{i,j\} \\
        b_{ij}+b_{ik}b_{kj} & b_{ik}>0,b_{kj}>0 \\
        b_{ij}-b_{ik}b_{kj} & b_{ik}<0,b_{kj}<0 \\
        b_{ij} & \text{otherwise}
    \end{array}\right. ,\, \, x_{k}^{\prime}=\frac{1}{x_{k}}\left[\prod_{i|b_{ik}>0}x_{i}^{b_{ik}}+\prod_{i|b_{ik}<0}x_{i}^{-b_{ik}}\right].
\end{equation}
The cluster algebra is generated by all the cluster variables (e.g. $\mathcal{A}$ coordinates) obtained through mutations. To calculate the relations between different $\mathcal{A}$ coordinates, we choose the following parameterization for $Z_{i=1,\ldots,7}$~\cite{He:2021eec},
\begin{equation}\label{eq:Zpara}
\begin{aligned}
    &Z_{1}=(1,0,0,0), \, Z_{2}=\left( \frac{1+f_{5}}{f_{5}},1,0,0 \right), \\ &Z_{3}=\left(\frac{f_1 f_2+f_1 f_4 f_2+f_1 f_3 f_4 f_2+f_2+1}{f_1 f_2 f_3 f_4 f_5},\frac{f_1 f_2+f_1 f_6 f_2+f_1 f_4 f_6 f_2+f_1 f_3 f_4 f_6 f_2+f_6 f_2+f_2+f_6+1}{f_1 f_2 f_3 f_4 f_6},1,0\right), \\
    &Z_{4}=\left(\frac{f_1 f_2+f_1 f_4 f_2+f_2+1}{f_1 f_2 f_3 f_4 f_5},\frac{\left(f_1 f_2+f_1 f_4 f_2+f_2+1\right) \left(f_6+1\right)}{f_1 f_2 f_3 f_4 f_6},f_1 f_2+f_1 f_4 f_2+f_2+1,1\right), \\
    &Z_{5}=\left(0,\frac{1}{f_3 f_6},f_2 \left(f_4 f_1+f_1+1\right),1\right), Z_{6}=\left(0,0,\frac{f_2 \left(f_1 f_2+f_1 f_4 f_2+f_2+1\right)}{f_2+1},1\right),Z_{7}=(0,0,0,1).
\end{aligned}
\end{equation}
and all symbol letters can be expressed as rational functions of polynomials over six $f_{i}$'s. For odd number $n$-particle kinematics, there is a unique way to uplift the mutable $\mathcal{A}$ coordinates to DCI cross ratios, which can be easily proved by setting an ansatz and show that the rank of $n$ linear equations is full. This is how \eqref{SalphabetAppx} is derived and we have colored $\mathcal{A}$ coordinates corresponding to mutable nodes blue as in \cite{Caron-Huot:2020bkp}. Then the initial quiver can be simplified by using $a_{ij}$ as cluster variables and suppress the frozen nodes (set to 1),
\begin{equation}\label{eq:initialquiver}
    \begin{tikzpicture}[>=stealth,boxed/.style={draw, rectangle}, every path/.style={shorten >=2pt, shorten <=2pt}]
    \matrix (m) [matrix of nodes,
        row sep=1.2em,
        column sep=2em,
        nodes={inner sep=1.5pt}] 
    {
        $a_{24}$   & $a_{37}$     \\
        $a_{13}$   & $a_{17}$     \\
        $a_{32}$   & $a_{27}$     \\
    };

    \draw[->] (m-1-1) -- (m-1-2);
    \draw[->] (m-1-2) -- (m-2-2);

    \draw[->] (m-1-1) -- (m-2-1);
    \draw[->] (m-1-2) -- (m-2-2);

    \draw[->] (m-2-1) -- (m-3-1);
    \draw[->] (m-2-1) -- (m-2-2);
    \draw[->] (m-2-2) -- (m-1-1);
    \draw[->] (m-2-2) -- (m-3-2);

    \draw[->] (m-3-1) -- (m-3-2);
    \draw[->] (m-3-2) -- (m-2-1);

\end{tikzpicture}
\end{equation}
Performing the mutation in the node order $a_{17},a_{27},a_{32},a_{37},a_{27},a_{37},a_{13}$~\footnote{We note that there are many different mutation orders which can generate the $E_{6}$ quiver. However, the final cluster algebra is the same since these different quivers are connected by a set of mutations.}, we will end up with the following quiver and cluster variables
\begin{equation}\label{eq:E6quiver}
    \begin{tikzpicture}[>=stealth,boxed/.style={draw, rectangle}, every path/.style={shorten >=2pt, shorten <=2pt}]
    \matrix (m) [matrix of nodes,
        row sep=1.2em,
        column sep=2em,
        nodes={inner sep=1.5pt}] 
    {
        &  & $a_{11}$ & &   \\
        $a_{33}$ & $a_{41}$ & $a_{62}$ & $a_{51}$ & $a_{24}$     \\
    };

    \draw[->] (m-1-3) -- (m-2-3);
    \draw[->] (m-2-1) -- (m-2-2);
    \draw[->] (m-2-2) -- (m-2-3);
    \draw[->] (m-2-4) -- (m-2-3);
    \draw[->] (m-2-5) -- (m-2-4);

\end{tikzpicture}
\end{equation}
This is a standard quiver for $E_{6}$ type cluster algebra. We can mutate the quiver \eqref{eq:E6quiver} while keeping its shape as $E_{6}$. It is interesting that $a_{1j}$ always appears in node $a_{11}$ and $a_{6j}$ always appears in node $a_{62}$. $a_{4j}$ will appear in node $a_{41}$ and $a_{51}$, and it is the same for $a_{5j}$. $a_{3j}$ will appear in node $a_{33}$ and $a_{24}$, and it is the same for $a_{2j}$. Thus, the heptagon symbol alphabet is described by above $E_{6}$ cluster algebra. And it is intriguing that the first entry set $\{a_{1j}\}$ and the last entry set $\{a_{2j},a_{3j}\}$ for our symbol bootstrap appear as the two ends of $E_{6}$ quiver.

The cluster algebra encodes not only information of symbol alphabets, but also how these symbol letters can appear in the symbol, which is called the cluster adjacency conjectures~\cite{Drummond:2017ssj}. It states that, for the BDS-like subtracted amplitudes, two distinct $\mathcal{A}$ coordinates (e.g. $a_{ij}$ in our example) can be adjacent in the symbol only if there is a cluster where they both appear. For example, in \eqref{eq:E6quiver}, $a_{11}$ and $a_{41}$ both appear, then $a_{11}$ \textit{can} be adjacent to $a_{41}$ in the symbol. On the contrary, $a_{11}$ and $a_{13}$ never appear in the same cluster, thus they are forbidden to be adjacent in the symbol. The cluster adjacency constraints are known to be equivalent to the extended Steinmann relations with the same first entries~\cite{Drummond:2017ssj,Caron-Huot:2020bkp}. In addition to the adjacency relations, the exchange relations of cluster algebra in \eqref{eq:exchangerels} impose further constraints on these possible adjacent pairs through the requirement of integrability. The integrability property, which guarantees that the symbol corresponds to some analytic function, requires that for any integrable symbol
\begin{equation}\label{eq:symboldef}
    \mathcal{S}(F^{(w)}) = \sum_{\alpha_{1},\ldots,\alpha_{w}} C_{\alpha_{1},\ldots,\alpha_{w}} [\mspace{1mu}\phi_{\alpha_{1}},\ldots, \phi_{\alpha_{w}}],
\end{equation}
it must satisfy the following constraints
\begin{equation}
\label{eq:integrability}
    \sum_{\alpha_{1},\ldots,\alpha_{w}} C_{\alpha_{1},\ldots,\alpha_{w}} \underbrace{[\mspace{1mu}\phi_{\alpha_{1}},\ldots, \phi_{\alpha_{w}}]}_{\text{omitting } \alpha_{j} , \alpha_{j{+}1}} \mathrm{d}\log \phi_{\alpha_{j}} \wedge \mathrm{d}\log \phi_{\alpha_{j{+}1}} = 0 \quad \forall j \in \{1,2,\ldots, w{-}1\}\,.
\end{equation}
Let us show an example how exchange relations impose further constraints through the requirement of integrability. Mutating the node $a_{11}$ in \eqref{eq:E6quiver}, we can derive from the exchange relation that
\begin{equation}
    a_{11}^{\prime}=\frac{1}{a_{11}}\left[1+a_{62}\right]=a_{13}. \,\, \Rightarrow \,\, \log a_{11}+\log a_{13} = \log (1+a_{62})
\end{equation}
It follows that
\begin{equation}
    \mathrm{d}\log a_{11}\wedge \mathrm{d}\log a_{62}+\mathrm{d}\log a_{13}\wedge \mathrm{d}\log a_{62} = \mathrm{d}\log (1+a_{62})\wedge \mathrm{d}\log a_{62} =0.
\end{equation}
Both $(a_{11},a_{62})$ and $(a_{13},a_{62})$ are allowed adjacency pairs. The exchange relation shows the possible way that they are combined to satisfy the integrability condition. Let us show another identity derived from initial quiver \eqref{eq:initialquiver} by mutating node $a_{24}$:
\begin{equation}
    a_{45}a_{24}=a_{17}+a_{13}a_{37}. \,\, \Rightarrow \,\, \mathrm{d}\log \frac{a_{45}a_{24}}{a_{17}}\wedge \mathrm{d}\log\frac{a_{13}a_{37}}{a_{17}}=\mathrm{d}\log\left(1+\frac{a_{13}a_{37}}{a_{17}}\right)\wedge \mathrm{d}\log\frac{a_{13}a_{37}}{a_{17}} =0.
\end{equation}
Especially, we give an identity that is related to the sequences we find in the main text:
\begin{equation}
    a_{11}a_{17}=a_{14}+a_{21}a_{37}. \,\, \Rightarrow \,\, \mathrm{d}\log \frac{a_{11}a_{17}}{a_{14}}\wedge \mathrm{d}\log\frac{a_{21}a_{37}}{a_{14}}=\mathrm{d}\log\left(1+\frac{a_{21}a_{37}}{a_{14}}\right)\wedge \mathrm{d}\log\frac{a_{21}a_{37}}{a_{14}} =0.
\end{equation}
However, such identities can be many and they may not be independent. We have checked that the exchange relations of $E_{6}$ cluster algebra can generate 132 independent $\mathrm{d}\log$ two form identities by walking through all 833 clusters. In practice, we can also solve all the relations between $861{=}42{\times} 41/2$ nontrivial $\mathrm{d}
\log\ldots\wedge\mathrm{d}
\log\ldots$ numerically using parameterization \eqref{eq:Zpara}. The independent relations obtained in this way agrees with what we derived from exchange relations. Thus the integrability conditions will give $729{=}861{-}132$ nontrivial constraints for each pair of adjacent slots in the symbol.

The BDS ansatz \cite{Bern:2005iz} captures the IR structure of the amplitudes. Thus, we can define the finite BDS-normalized MHV and NMHV amplitude as
\begin{equation}
    \mathcal{B}_n = \frac{\mathcal{A}_n^{\text{MHV}}}{\mathcal{A}_n^{\text{BDS}}} = \exp [R_n] \, , \quad B_n = \frac{\mathcal{A}_n^{\text{NMHV}}}{\mathcal{A}_n^{\text{BDS}}} = \frac{\mathcal{A}_n^{\text{NMHV}}}{\mathcal{A}_n^{\text{MHV}}} \frac{\mathcal{A}_n^{\text{MHV}}}{\mathcal{A}_n^{\text{BDS}}} = \mathcal{P}_n \mathcal{B}_n \, ,
\end{equation}
where $\mathcal{A}_n^{\text{BDS}}$ is the BDS ansatz for the superamplitude. In the collinear limit $n \parallel (n-1)$, where momenta of two neighboring particles $n$ and $(n-1)$ are taken to be proportional (\textit{e.g.} the explicit parameterization in momentum twistor for $n=7$ is given in \eqref{eq:momtwicoll}), the BDS-normalized amplitude smoothly reduces to the same quantity with one fewer particle: $\lim_{n \parallel (n-1)} R_n = R_{n-1}$ and $\lim_{n \parallel (n-1)} B_n = B_{n-1}$. However, $R_n$ and $B_n$ do not satisfy the (extended) Steinmann condition (or the cluster adjacency condition), since the BDS-normalized amplitudes depend on three-particle Mandelstam invariants. In order to impose the cluster adjacency condition, we should consider the BDS-like ansatz:
\begin{equation}
    \mathcal{A}_n^{\text{BDS-like}} = \mathcal{A}_n^{\text{BDS}} \exp \left( \frac{\Gamma_{\text{cusp}}}{4} Y_n \right),
\end{equation}
where $\Gamma_{\text{cusp}}$ is the cusp anomalous dimension, which can be set to $4g^2$ after omitting the constants with higher transcendental weight, since we only consider the symbol of the amplitudes; and $Y_n$ is a dual conformally invariant weight-2 function containing all the dependence on the three-particle invariants and for $n=7$ we have $ S[Y_7] = [a_{11}, \frac{a_{21} a_{37}}{a_{25}}] + {\rm dih.} $ The BDS-like-normalized MHV and NMHV amplitudes are then defined as
\begin{equation}
    \mathcal{E}_{n,0} = \frac{\mathcal{A}_n^{\text{MHV}}}{\mathcal{A}_n^{\text{BDS-like}}} = \exp \left( R_n - \frac{\Gamma_{\text{cusp}}}{4} Y_n \right) \, , \quad \mathcal{E}_{n,1} = \frac{\mathcal{A}_n^{\text{NMHV}}}{\mathcal{A}_n^{\text{BDS-like}}} = \mathcal{P}_n \mathcal{E}_{n,0}.
\end{equation}
It is easy to see that $\mathcal{E}_{7}^{(1)}=-Y_{7}$ since $R_n^{(1)}=0$. And for MHV we give the explicit formula for each loop order in the following (the subscripts denoting MHV are suppressed for simplicity)
\begin{equation}\label{eq:EtoR}
\begin{aligned}
    &\mathcal{E}_{7}^{(2)} = R_{7}^{(2)}+\frac{1}{2}(\mathcal{E}_{7}^{(1)})^{2},\\
    &\mathcal{E}_{7}^{(3)} = R_{7}^{(3)}+R_{7}^{(2)}\mathcal{E}_{7}^{(1)}+\frac{1}{6}(\mathcal{E}_{7}^{(1)})^{3}, \\
    &\mathcal{E}_{7}^{(4)} = R_{7}^{(4)}+R_{7}^{(3)}\mathcal{E}_{7}^{(1)}+\frac{1}{2}(R_{7}^{(2)})^{2}+\frac{1}{2}R_{7}^{(2)}(\mathcal{E}_{7}^{(1)})^{2}+\frac{1}{24}(\mathcal{E}_{7}^{(1)})^{4}, \\
    &\mathcal{E}_{7}^{(5)} = R_{7}^{(5)}+R_{7}^{(4)}\mathcal{E}_{7}^{(1)}+R_{7}^{(2)}R_{7}^{(3)}+\frac{1}{2}R_{7}^{(3)}(\mathcal{E}_{7}^{(1)})^{2}+\frac{1}{2}(R_{7}^{(2)})^{2}\mathcal{E}_{7}^{(1)}+\frac{1}{6}R_{7}^{(2)}(\mathcal{E}_{7}^{(1)})^{3}+\frac{1}{120}(\mathcal{E}_{7}^{(1)})^{5} \, .
\end{aligned}
\end{equation}

As mentioned in the main text, we also impose the symmetry conditions for BDS-like-normalized amplitude. The dihedral transformation for the $E_6$ symbol alphabet is defined as
\begin{equation}
\begin{aligned}\label{eq:dihedrala}
    \text{Cyclic transformation:}& \quad a_{li}\to a_{l,i+1}\,,\\
    \text{Flip transformation:}& \begin{cases}
    a_{2i}\leftrightarrow a_{3,8-i}\\
    a_{li}\to a_{l,8-i}&\text{for}\quad l\ne 2,3\,.
    \end{cases}
\end{aligned}
\end{equation}
Under these transformations, the symbol of MHV heptagon $\mathcal{S}^{(L)}$ and the NMHV $S[E_{0}^{(L)}]$ are invariant. Furthermore, $\mathcal{S}^{(L)}$ is also invariant under the parity transformation
\begin{equation}
    \text{Parity transformation:}\quad a_{2,i+1} \leftrightarrow a_{3,i}, \qquad a_{4i} \leftrightarrow a_{5i},
\end{equation}
since the MHV and $\overline{\text{MHV}}$ amplitudes differ only in their tree-level prefactors. For NMHV $E_{12}$ and $E_{14}$, they are invariant under the flip $Z_i \rightarrow Z_{3-i}$ and $Z_i \rightarrow Z_{5-i}$ respectively, which is inherited by the corresponding $e_{ij}$'s.

\section{More details on the computation and the results}
In this section, we briefly summarize the algorithm we employed, the specific computation procedure, and the results obtained.
\paragraph{Algorithm} Our algorithm is based on Appendix B of \cite{Dixon:2016nkn}. For a weight-$w$ integrable symbol defined in \eqref{eq:symboldef}, the integrability condition (with repeated indices indicating tensor contraction) reads
\begin{equation}
    C_{\alpha_1,\dots,\alpha_j,\alpha_{j+1},\dots,\alpha_w} \mspace{2mu} \mathrm{d}\log\phi_{\alpha_j}\wedge \mathrm{d}\log\phi_{\alpha_{j+1}}=0.
\end{equation}

Choosing a basis of $729$ independent $\mathrm{d}\log$ 2-forms (denoted by $(\mathrm{d}\log\ldots\wedge \mathrm{d}\log\ldots)_k$), we expand every wedge $\mathrm{d}\log\phi_{i}\wedge \mathrm{d}\log\phi_{j}$ as a linear combination $\mathrm{d}\log\phi_{i}\wedge \mathrm{d}\log\phi_{j}=\sum_{k=1}^{729} W_{ijk}(\mathrm{d}\log\ldots\wedge \mathrm{d}\log\ldots)_k$. The integrability condition then becomes the following tensor equations:
\begin{equation}
    C_{\alpha_1,\dots,\alpha_i,\alpha_{i+1},\dots,\alpha_w} W_{\alpha_i,\alpha_{i+1},k}=0\ ,\quad \forall k\in\{1,\dots,729\}.
\end{equation}

In addition, there are $42^2-840=924$ forbidden adjacent pairs for heptagon symbols. Denote the $k$-th forbidden pair by $(\phi_{\beta_k},\phi_{\gamma_k})$, the cluster adjacency conditions can then be written as
\begin{equation}
    C_{\alpha_1,\dots,\alpha_i,\alpha_{i+1},\dots,\alpha_w} V_{\alpha_i,\alpha_{i+1},k}=0\ ,\quad V_{ijk}=\delta_{i,\beta_k}\delta_{j,\gamma_k},\quad \forall k\in\{1,\dots,924\}.
\end{equation}

Combining $V_{ijk}$ and $W_{ijk}$ together and eliminating $462$ linearly dependent ones yields $924+729-462=1191$ independent equations on adjacent entries of heptagon symbols. We arrange these equations into a rank-$3$ tensor of dimension $(42,42,1191)$ which we denote by $\texttt{dlogmat}_{ijk}$. The constraints on heptagon symbols can be therefore compactly expressed as
\begin{equation}\label{eq:dlogmat_cond}
    C_{\alpha_1,\dots,\alpha_i,\alpha_{i+1},\dots,\alpha_w}\mspace{2mu} \texttt{dlogmat}_{\alpha_i,\alpha_{i+1},k}=0.
\end{equation}

For high weight symbol, a full rank-$w$ tensor is always too large to bootstrap. Instead, we expand the weight-$w$ symbol basis into tensor products of weight-$(w-1)$ symbol basis and letters, then solve for the expansion coefficients, which encode the $\Delta_{w-1,1}$ or $\Delta_{1,w-1}$ coproduct component of the symbols. That is to say, let $C^{k}_{\alpha_1,\dots,\alpha_w}$ be the $k$-th basis of the weight-$w$ symbols, we introduce the following `forward expansion coefficients' $F^{(w)}_{i,\alpha_w,j}$ encoding the $\Delta_{w-1,1}$ coproduct entry of $C^{j}_{\alpha_1,\dots,\alpha_w}$ w.r.t. $\alpha_w$, as well as the `backward expansion coefficients' $L^{(w)}_{j,\alpha_1,i}$ encoding the $\Delta_{1,w-1}$ coproduct entry of $C^{j}_{\alpha_1,\dots,\alpha_w}$ w.r.t. $\alpha_1$, such that
\begin{equation}
    C^{i}_{\alpha_1,\dots,\alpha_{w-1}}F^{(w)}_{i,\alpha_w,j} =C^{j}_{\alpha_1,\dots,\alpha_w} =L^{(w)}_{j,\alpha_1,i}\, C^{i}_{\alpha_2,\dots,\alpha_w}\ ,\quad w>1
\end{equation}
and the first- and last- entry conditions are imposed here by defining
\begin{equation}
    F^{(1)}_{\alpha_1,i}:=\begin{cases}
        \delta_{\alpha_1,i} & \alpha_1 \in \text{first entries}\\
        0 & \alpha_1 \notin \text{first entries}
    \end{cases}\ ,\quad
    L^{(1)}_{j,\alpha_w}:=\begin{cases}
        \delta_{j,\alpha_w} & \alpha_w \in \text{last entries}\\
        0 & \alpha_w \notin \text{last entries}
    \end{cases}\ ,
\end{equation}

Hence the full symbol can be written as the product
\begin{equation}
    C^{k}_{\alpha_1,\dots,\alpha_w} = F^{(1)}_{\alpha_1,i_1} F^{(2)}_{i_1,\alpha_2,i_2}\cdots F^{(w-1)}_{i_{w-2},\alpha_{w-1},i_{w-1}} F^{(w)}_{i_{w-1},\alpha_w,k}
    = L^{(w)}_{k,\alpha_1,j_{w-1}} L^{(w-1)}_{j_{w-1},\alpha_2,j_{w-2}}\cdots L^{(2)}_{j_2,\alpha_{w-1},j_1} L^{(1)}_{j_1,\alpha_w}.
\end{equation}

And the \texttt{dlogmat} conditions~\eqref{eq:dlogmat_cond} now reduce to constraints on sequential coproduct entries:
\begin{equation}\label{eq:FCC_LCC_cond}
    F^{(w-1)}_{i_{w-2},\alpha_{w-1},i_{w-1}} F^{(w)}_{i_{w-1},\alpha_w,i_{w}} \texttt{dlogmat}_{\alpha_{w-1},\alpha_w,k}=0\ ,\quad L^{(w)}_{j_w,\alpha_1,j_{w-1}} L^{(w-1)}_{j_{w-1},\alpha_2,j_{w-2}} \texttt{dlogmat}_{\alpha_w,\alpha_{w-1},k}=0\ ,\quad \forall\,w\,,
\end{equation}
which allows us to determine $F^{(w)}$ and $L^{(w)}$ from $F^{(w-1)}$ and $L^{(w-1)}$, and therefore iteratively build up the solution space starting from the known $F^{(1)}$ and $L^{(1)}$ fixed by the first/last-entry conditions. Note that for NMHV cases, we should apply only the cluster adjacency part of $\texttt{dlogmat}_{ijk}$ to the last two entries of $e_{ij}$~\eqref{eq:eij_coproduct}, and impose the integrability of the last two entries only on $E_{ij}$ summing from $e_{ij}$~\eqref{eq:eij_to_Eij}.

To obtain the solution space at weight $2L$ that satisfies both first- and last-entry conditions, rather than imposing the last-entry condition directly on $F^{(2L)}$ (or vice versa), we ``sew'' the forward and backward expansions using sewing matrices $M_{i_w,j_{2L-w}}$~\cite{Basso:2024hlx}. Specifically, we solve for matrices $M^{k}_{i_w,j_{2L-w}}$ satisfying:
\begin{equation}\label{eq:sew_cond}
    F^{(w)}_{i_{w-1},\alpha_w,i_w} M^{k}_{i_w,j_{2L-w}} L^{(2L-w)}_{j_{2L-w},\alpha_{w+1},j_{2L-w-1}} \texttt{dlogmat}_{\alpha_w,\alpha_{w+1},k}=0\,,
\end{equation}
where each matrix $M^{k}$ is a solution of the above equation. The specific choice of $w$ does not affect the final result, though $w=2L-1$ or $2L-2$ usually leads to smaller-scale computations. The full solution symbols of integrability, cluster adjacency, first entry and last entry conditions are then given by
\begin{equation}
    C^{k}_{\alpha_1,\dots,\alpha_{2L}}=F^{(1)}_{\alpha_1,i_1} \cdots  F^{(w)}_{i_{w-1},\alpha_w,i_w} M^{k}_{i_w,j_{2L-w}} L^{(2L-w)}_{j_{2L-w},\alpha_{w+1},j_{2L-w-1}} \cdots L^{(1)}_{j_1,\alpha_{2L}}.
\end{equation}

Apart from the above conditions acting on adjacent entries, the conditions with some transformation $U$ acting on each letter can also be conveniently solved based on the sequential coproduct entries and the sewing matrix. Consider conditions in the following form:
\begin{equation}\label{eq:U_cond}
    U_{\beta_1,\alpha_1} \cdots U_{\beta_{2L},\alpha_{2L}} C_{\alpha_1,\dots,\alpha_{2L}} = \tilde{C}_{\beta_1,\dots,\beta_{2L}},
\end{equation}
where the target symbol $\tilde{C}_{\beta_1,\dots,\beta_{2L}}$ can be $C_{\alpha_1,\dots,\alpha_{2L}}$ itself ({\it e.g.} when $U$ represents the symmetry of the symbol), or some other symbol ({\it e.g.} when $U$ is a projection to lower-point symbol when considering the collinear limit). Instead of solving \eqref{eq:U_cond} directly, it will be helpful to introduce the following induced transformation matrix $T$ satisfying:
\begin{equation}
\begin{array}{ll}
    U_{\beta_w,\alpha_w} T^{(w-1),F}_{i'_{w-1},i_{w-1}} F^{(w)}_{i_{w-1},\alpha_w,i_w} = \tilde{F}^{(w)}_{i'_{w-1},\beta_w,i'_w} T^{(w),F}_{i'_{w},i_{w}}\ , & U_{\beta_1,\alpha_1} F^{(1)}_{\alpha_1,i_1} = \tilde{F}^{(1)}_{\beta_1,i'_1} T^{(1),F}_{i'_1,i_1} \\
   U_{\beta_{2L-w+1},\alpha_{2L-w+1}} L^{(w)}_{j_w,\alpha_{2L-w+1},j_{w-1}} T^{(w-1),L}_{j_{w-1},j'_{w-1}} = T^{(w),L}_{j_{w},j'_{w}} \tilde{L}^{(w)}_{j'_w,\beta_{2L-w+1},j'_{w-1}} \ , & U_{\beta_{2L},\alpha_{2L}} L^{(1)}_{j_1,\alpha_{2L}} = T^{(1),L}_{j_1,j'_1} \tilde{L}^{(1)}_{j'_1,\beta_{2L}}
\end{array},
\end{equation}
where $\tilde{F},\tilde{L},\tilde{M}$ denote the corresponding objects of $\tilde{C}_{\beta_1,\dots,\beta_{2L}}$. The condition~\eqref{eq:U_cond} then reduces to
\begin{equation}\label{eq:T_cond}
    T^{(w),F}_{i'_w,i_w} M_{i_w,j_{2L-w}} T^{(2L-w),L}_{j_{2L-w},j'_{2L-w}} = \tilde{M}_{i'_w,j'_{2L-w}} 
\end{equation}

The dihedral, parity, $\mathbb{Z}_2$ symmetry conditions and the collinear limit can all be implemented using \eqref{eq:U_cond} or \eqref{eq:T_cond}, by imposing $C_{\alpha_1,\dots,\alpha_{2L}}=c_k C^k_{\alpha_1,\dots,\alpha_{2L}}$ or $M_{i_w,j_{2L-w}}=c_k M^k_{i_w,j_{2L-w}}$ and solve for the coefficient $c_k$'s. For symmetry conditions, $\tilde{C}$ is just $C$ itself; and for collinear limit, $\tilde{C}$ is six-point BDS-like amplitudes along with divergent terms containing $\eta^{\prime},\tau^{\prime}$. In the collinear case, we actually only employ the divergent terms in $\tilde{C}$, which can already fix all $c_k$'s.

\paragraph{Computations} The linear equations in the above algorithm can be efficiently solved using \texttt{SparseRREF}~\cite{SparseRREF}, a \texttt{C++} library designed for computing the row-reduced echelon form (RREF) of sparse matrices, which can use the \textsc{Mathematica} \texttt{SparseArray} objects in \texttt{WXF} format for input and output. In addition to its core RREF capability, \texttt{SparseRREF} supports various operations for sparse tensors and matrices including addition, multiplication (tensor contraction), transposition, reshaping, and others. One of the authors of this paper is a contributor to the library.

Following the instructions in \texttt{README.md} and \texttt{math\_link.cpp}, \texttt{SparseRREF} can be compiled as a dynamic library and loaded as functions in \textsc{Mathematica}. With this setup, computations up to four loops can be carried out on a modern laptop. The five-loop computation, nevertheless, demands resources orders of magnitude beyond the prior cases, prompting the development of a standalone \texttt{C++} program. This program utilizes modified functions from the \texttt{SparseRREF} library, and was executed on the high-performance computing cluster of ITP-CAS bypassing \textsc{Mathematica}. To demonstrate, the MHV bootstrap took approximately 24 hours on a node equipped with 160-cores Intel Xeon E7-8870 @ 2.10 GHz and reached a peak memory usage of about 1.3 TB.

\paragraph{Results} We provide the result of heptagon five-loops symbols as an ancillary file and can be found in \cite{Symbol_Data}. In particular, we record the statistics of all coefficients at five-loop for both MHV and NMHV symbols in the ancillary file \texttt{heptagon\_coefficient\_counting.m} along with the  arXiv submission of this paper. To use this file, just load it in \textsc{Mathematica}, then the coefficient countings for the MHV amplitude and the NMHV $E_0$, $E_{12}$, $E_{14}$ functions are respectively given by variables \texttt{mhvL5counting}, \texttt{nmhvE0L5counting}, \texttt{nmhvE12L5counting}, and \texttt{nmhvE14L5counting}, in the form of a list of \texttt{\{coefficient, number of terms with that coefficient\}}. It exhibits some interesting feature similar to those found for the coefficients of $f$ graphs~\cite{He:2024cej,Bourjaily:2025iad}: for example, most coefficients are concentrated around zero, {\it e.g.} the majority of coefficients are $\pm 1$, and large coefficients (in magnitude) are very rare with some huge gaps between them.  We illustrate the distribution of coefficients for $L=5$ in the following Figures. 

\begin{figure}[H]
    \centering
    \includegraphics[scale=0.7]{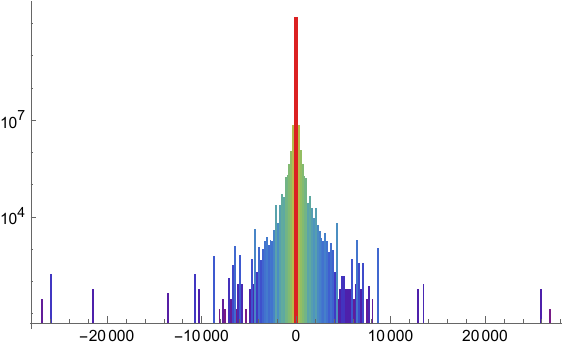}
    \caption{Coefficient distribution of five-loop MHV symbol.}
    \label{fig:coeff_mhv_w10_dist}
\end{figure}

\begin{figure}[H]
	\centering
	\subfloat[$E_0$]{\label{fig:coeff_nmhv_E0_w10_dist} \includegraphics[width=0.31\linewidth]{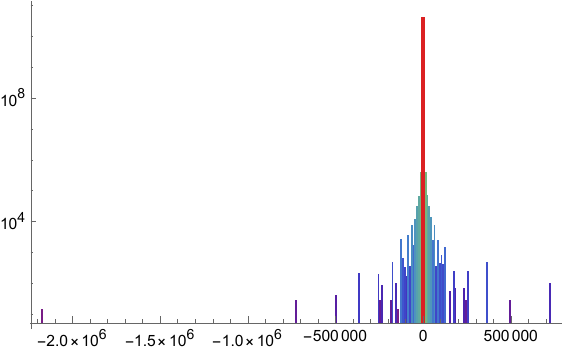}}
	\subfloat[$E_{12}$]{\label{fig:coeff_nmhv_E12_w10_dist} \includegraphics[width=0.31\linewidth]{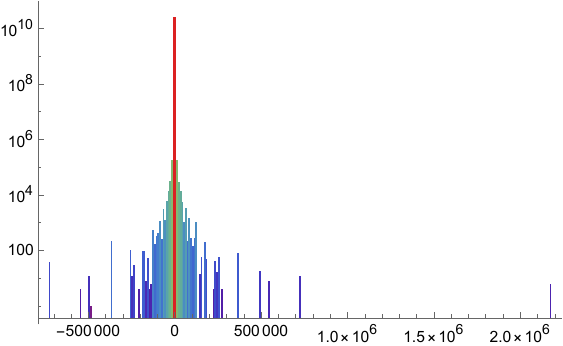}}
	\subfloat[$E_{14}$]{\label{fig:coeff_nmhv_E14_w10_dist} \includegraphics[width=0.31\linewidth]{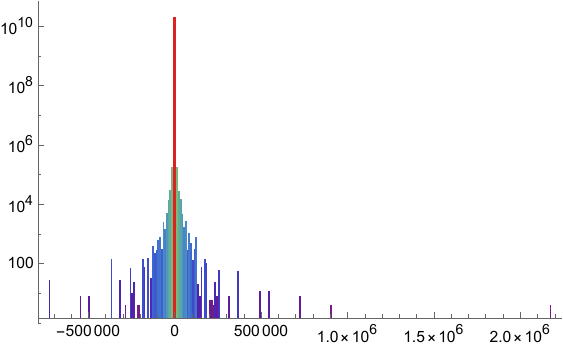}}
	\caption{Coefficient distribution of five-loop NMHV symbol.}
	\label{fig:coeff_nmhv_w10_dist}
\end{figure}

\end{document}